\documentclass[twocolumn,prl]{revtex4}
\usepackage{epsfig}
\begin{document}

\title{Intrinsic self-rotation of BEC Bloch wave packets}
%
\author{Roberto B.  Diener}
\affiliation{Department of Physics, The University of Texas,
Austin, Texas 78712-1081}
\author{Artem M. Dudarev}
\affiliation{Department of Physics, The University of Texas,
Austin, Texas 78712-1081}
\affiliation{Center for  Nonlinear
Dynamics, The University of Texas, Austin, Texas 78712-1081}
\author{Ganesh Sundaram}
\affiliation{Department of Physics, The University of Texas,
Austin, Texas 78712-1081}
\author{Qian Niu}
\affiliation{Department of Physics, The University of Texas,
Austin, Texas 78712-1081}

\begin{abstract}
The semiclassical theory of Bloch wave packet dynamics predicts a self-rotation
angular momentum in asymmetric periodic potentials, which has never been 
observed. We show how this is manifested in
 Bose-Einstein condensed atoms in optical
lattices. 
Displacing the wave
packet to a corner of the Brillouin zone we obtain a current
distribution with a non-quantized angular momentum, independent of
the size of the distribution. A weak interatomic interaction does
not modify the results, affecting only the rate of spreading in
the lattice. A strong repulsive interaction results in a collapse of
the wavefunction into matter-wave lattice solitons.
\end{abstract}

\maketitle

The semiclassical equations of motion for electrons in 
crystals have played a fundamental role in the 
theory of charge
transport in metals and semiconductors~\cite{Ashcroft-Mermin}. 
Because of Berry phase and gradient corrections~\cite{Chang-Niu,Sundaram-Niu},
two striking effects occur for systems without time reversal or
spatial inversion symmetry. 
The
first one is a lateral displacement of the particles
relative to the external force, and is responsible for the anomalous 
Hall effect in ferromagnets~\cite{Jungwirth,Dimi}.
The second one is
the appearance of an orbital magnetization, due to the rotation of
the wave packet around its center.
This latter effect has received little attention and 
no experimental observation of it has been performed so far.
In this letter we propose a
demonstration of this wave packet self-rotation using
ultracold atoms in optical lattices. Such a system has been employed in the
past to exhibit a number of basic phenomena in condensed matter 
physics~\cite{Bloch,WSL}.

A dual purpose of this work is to study the generation of
novel matter-wave states using the knowledge gathered in electron transport
theory. Bose-Einstein condensates can be manipulated experimentally 
with great precision~\cite{Pethick-Smith} and the
equation of motion for the condensate wavefunction in a periodic potential 
coincides with the one for electrons moving in a crystal. 
Several experimental
studies of condensates in optical lattices have already been 
performed~\cite{Kasevich2,Mott}. The
generation of states in an asymmetrical potential must take into 
consideration the two effects mentioned before.


The semiclassical description assumes that the wave function for a
particle is a superposition of Bloch waves from a single band
\begin{equation}\label{expansion}
| \Psi \rangle = \int_{BZ} d{\bf k} \,f({\bf k})\, |\Psi_{\bf k} \rangle,
\end{equation}
in which the distribution $f(\bf k)$ is centered at a point ${\bf
k}_c$ in the Brillouin zone with a small dispersion $\sigma_k$.
This in turn yields a spatial dispersion $\sigma_r$ much larger
than the size of a unit cell. The Bloch waves have the
property that $|\Psi_{\bf k} \rangle = e^{i{\bf k}\cdot{\bf r}}
|u_{\bf k} \rangle$, where $u_{\bf k}({\bf r})$ preserves the
periodicity of the lattice potential $V_{\rm latt}({\bf r})$.
These wave functions are the solutions of the equation
\begin{eqnarray}
\hat{H}_0({\bf k}) |u_n({\bf k})\rangle &\equiv& [{1\over 2 M} ({\bf
p} + \hbar {\bf k})^2 + V_{\rm latt}({\bf r})]  |u_n({\bf k})\rangle\nonumber \\
&=& {\cal E}_n({\bf k}) |u_n({\bf k})\rangle.
\end{eqnarray}
We restrict our study to a single band, thus dropping the band
index $n$ in what follows.

By defining the Lagrangian ${\cal L}({\bf k}_c, {\bf r}_c,
\dot{\bf k}_c, \dot{\bf r}_c) = \langle \Psi |(i\hbar {\partial
\over
\partial t} - \hat{H}) |\Psi \rangle$ using as coordinates the position of
the center of the wave packet in real and reciprocal space, we can
find the equations of motion, which read
\begin{eqnarray}
\dot{\bf r}_c&=&{1\over \hbar}{\partial {\cal E}_S({\bf k}_c)
\over \partial {\bf k}_c}-\dot{\bf k}_c
\times {\bf \Omega}({\bf k}_c),\label{semiclassical1}\\
\hbar\dot{\bf k}_c&=&-e{\bf E}({\bf r}_c) - e \dot{\bf r}_c \times {\bf
B}({\bf r}_c).\label{semiclassical2}
\end{eqnarray}
For atoms in an optical lattice, the ``electric" and ``magnetic"
forces are the inertial Coriolis forces, if we
describe the motion in a frame moving with the lattice which is
linearly accelerated and/or rotated, respectively. Equation
(\ref{semiclassical2}) is the Lorenz force equation found in any
solid-state physics textbook, but (\ref{semiclassical1}) contains
two corrections from the standard result. One is the presence of
the Berry curvature, defined as
\begin{equation}
{\bf \Omega}({\bf k}_c) = i \langle {\partial u \over \partial
{\bf k}_c} | \times | {\partial u \over \partial {\bf k}_c}
\rangle
\end{equation}
(notice that this can be nonzero for complex $u({\bf r})$). It
acts dynamically as the symmetrical analogue of the magnetic field
if we exchange the ${\bf r}_c$ and ${\bf k}_c$ variables, and is
related to the Hall effect in magnetic sub-bands~\cite{Chang-Niu} 
and the anomalous Hall effect in ferromagnets~\cite{Jungwirth,Dimi}.
The second correction is to the band energy, which contains a term
proportional to the orbital magnetization energy, ${\cal E}_S({\bf
k}_c) = {\cal E}({\bf k}_c) + {e\over 2M} {\bf S}({\bf k}_c) \cdot {\bf
B}({\bf r}_c)$. This magnetization is proportional to the internal
angular momentum of the wave packet, i.e.
\begin{equation}
{\bf S} = \int d{\bf r} \, ({\bf r} -{\bf r}_c) \times {\bf
J}({\bf r}),
\end{equation}
where ${\bf J}({\bf r})$ is the current density~\cite{Chang-Niu}.
In terms of the Bloch waves, it is given by
\begin{equation}
{\bf S}({\bf k}_c) = i {M \over \hbar} \langle {\partial u \over
\partial {\bf k}_c} | \times (H({\bf k}_c) - {\cal E} ({\bf k}_c))
| {\partial u \over
\partial {\bf k}_c} \rangle.
\end{equation}
As long as the semiclassical approximation holds, 
this result is independent of the distribution $f({\bf k})$ used,
and hence of the width of the wave packet in real space. In
particular it is independent of time for a fixed value of ${\bf
k}_c$; the wave packet maintains its angular momentum constant as
it spreads on the lattice.

The reason that in most circumstances these corrections to the
semiclassical equations of motion do not need to be taken into
consideration is the constraints imposed on them by the symmetry
of the Hamiltonian. If the system possesses time-reversal
symmetry, then both vectors ${\bf \Omega}$ and ${\bf S}$ are odd
functions of ${\bf k}_c$. On the other hand, if it possesses
inversion symmetry then they are even functions of ${\bf k}_c$.
This implies that in systems with both symmetries they vanish
throughout the Brillouin zone.

\begin{figure}
\centering \epsfig{file=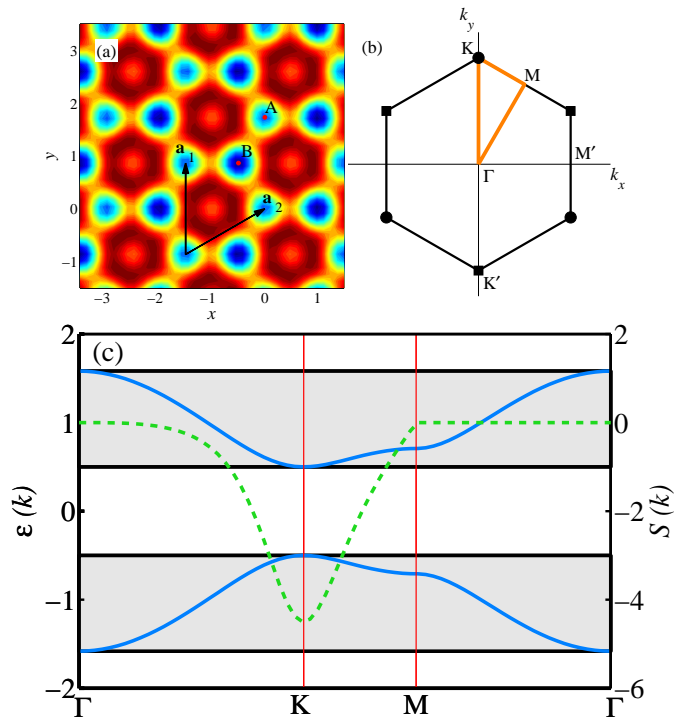, width=3.5in}
\caption{(a) potential energy seen by the atoms; the blue color corresponds to
the lowest values. In the tight-binding
approximation, particles are localized in the A and B sites only. 
The corresponding Brillouin zone in reciprocal space is shown in (b);
due to symmetry, all points marked by a circle correspond to the
same K point.
In (c) we show the band structure (solid line, left axis) and the
angular momentum $S$ (dashed line, right axis) for a tight binding band with
$\epsilon_g = h =1$.
} \label{figure1.fig}
\end{figure}

We will consider a system of Bose-condensed atoms moving in an
asymmetric two-dimensional optical lattice.
The interaction between the atoms is neglected for the moment.
In order to break
inversion symmetry, we consider an asymmetric hexagonal
lattice, as shown in Fig. \ref{figure1.fig}(a). 
In the
tight-binding regime, in which the potential wells denoted by blue
color are deep, the atoms are located in the ground state of these
wells. Since the calculations can be performed analytically in
this case, we consider it first~\cite{GaneshThesis}.


Let the on-site energies for sites A and B be $\epsilon_A =
\epsilon_g/2$ and $\epsilon_B = -\epsilon_g/2$, and the tunnelling
matrix element be non-vanishing only between nearest neighbors
with magnitude $h$. The lattice is triangular, with basis vectors
${\bf a}_1 = \sqrt{3}a {\bf e}_y$ and ${\bf a}_2 = {3\over 2} a
{\bf e}_x + {\sqrt{3}\over 2} a {\bf e}_y$; here $a$ is the
distance between nearest neighbors.  Numbering the sites in the
lattice by $(m,n)$ such that ${\bf R}_{m,n} = m{\bf a}_1 + n {\bf
a}_2$, we obtain the discrete Schr\"odinger equation for the
eigenvectors of the Hamiltonian
\begin{eqnarray}
{\cal E} \,\psi^A_{m,n} &=& \epsilon_A \,\psi^A_{m,n} - h
(\psi^B_{m,n+1} +\psi^B_{m,n}+\psi^B_{m+1,n}) \label{psiA},\\
{\cal E} \,\psi^B_{m,n} &=& \epsilon_B \,\psi^B_{m,n} - h
(\psi^A_{m,n-1} +\psi^A_{m,n}+\psi^A_{m-1,n}) \label{psiB}.
\end{eqnarray}

The Bloch waves for this Hamiltonian are calculated by writing
$\psi^{A,B}_{m,n} = \phi^{A,B} \, \exp(i{\bf k}\cdot {\bf R}_{m,n})$,
with which we obtain
\begin{equation}
H({\bf k}) =
\left(
\begin{array}{cc}
\epsilon_A & V({\bf k}) \\
V({\bf k})^* & \epsilon_B
\end{array}\right).
\end{equation}
The off-diagonal element is $V({\bf k}) = -h (1+e^{i{\bf k}\cdot
{\bf a}_1} + e^{i{\bf k}\cdot {\bf a}_2})$. The diagonalization of
the Bloch Hamiltonian is straightforward, yielding the two energy
bands
\begin{equation}
{\cal E}_\pm ({\bf k}) = \pm \sqrt{\left ({\epsilon_g\over 2} \right )^2 +
|V({\bf k})|^2}.
\end{equation}
The energy bands are depicted in Fig. 
\ref{figure1.fig}(c). The bottom band has a
minimum at the origin ($\Gamma$ point) and a maximum at the K
point, ${\bf k}_0 = 4\pi/(3\sqrt{3}a)\, {\bf e}_y$, at which
$V({\bf k}_0) = 0$. This is also the point of closest approach to
the excited band, with the gap being equal to $\epsilon_g$.

For the ground state wave function,
\begin{equation}
{\phi^A \over \phi^B} = -{V({\bf k}) \over {\cal E}_+ +
(\epsilon_g/2)}.
\end{equation}
In two dimensional systems, the Berry curvature and the angular
momentum are directed along the perpendicular, $z$-direction.
Their values for the system are
\begin{eqnarray}
\Omega ({\bf k}) &=& {\sqrt{3} \epsilon_g h^2 a^2 \over 16 {\cal
E}_+^3} [\sin(\sqrt{3}k_ya)-2\sin(\sqrt{3}k_ya/2)\cos(3k_xa/2)]
 \\
S ({\bf k})&=& {M\over \hbar} (2{\cal E}_+) \Omega({\bf
k}).\label{Mk}
\end{eqnarray}
An intriguing relation at the K point, where $S$ is minimum, is
\begin{equation}
S({\bf k}_0)= \hbar {M\over M^*}
\end{equation}
where $M^*$ is the effective mass, given by $M^* = -2\epsilon_g
\hbar^2 /(9h^2a^2)$.

The angular momentum $S$ is plotted in reciprocal space in figure
\ref{figure1.fig}(c) (dashed green line). From (\ref{Mk}) we see
that the dependence of the Berry curvature on quasimomentum in
reciprocal space is very similar. In particular, both of these
quantities attain their maximum absolute value at ${\bf k}_0$ and
all symmetrically located points. The value attained diverges as
the gap size $\epsilon_g$ goes to zero. Notice that the angular
momentum is not quantized.

\begin{figure}
\centering \epsfig{file=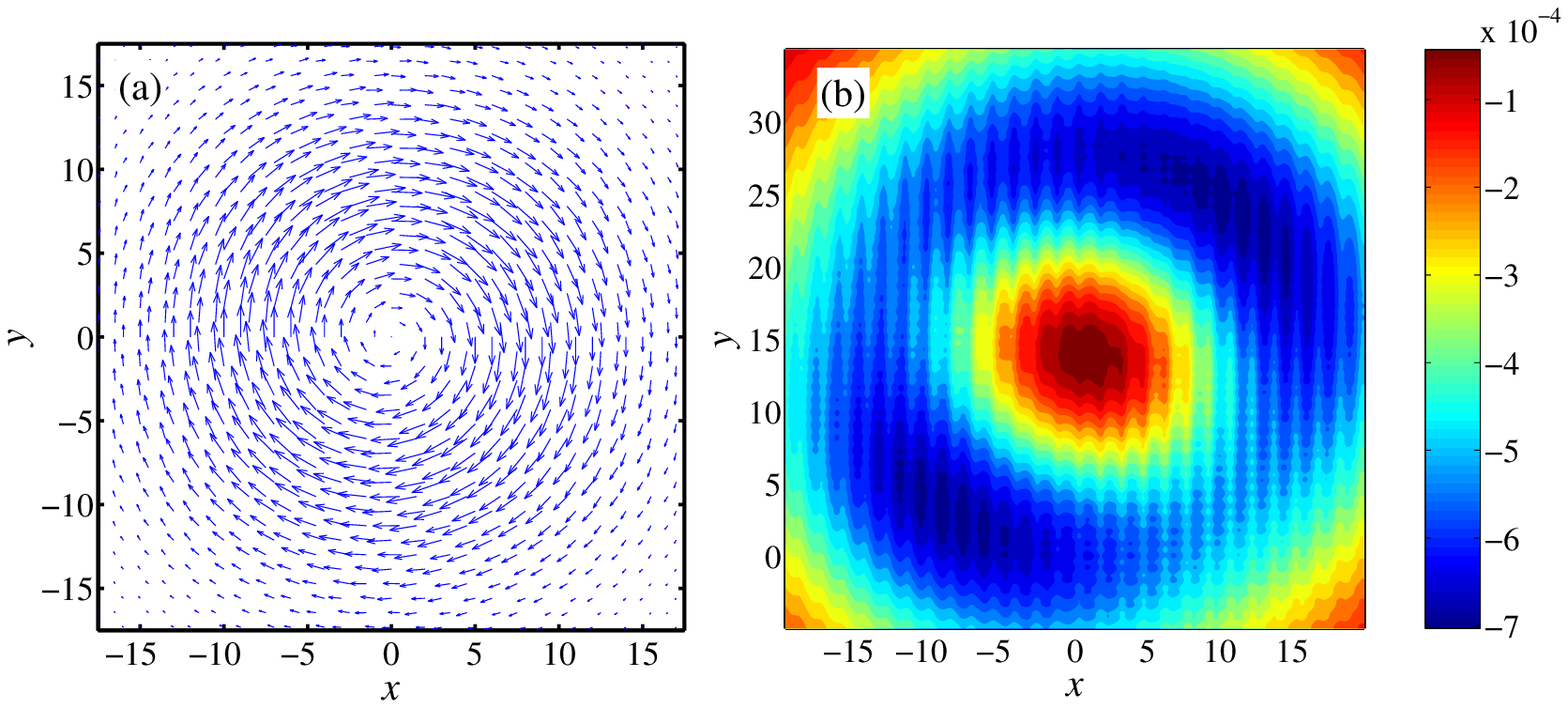, width=3.5in} \caption{(a): 
distribution of currents in the tight-binding limit for a
wave packet centered at ${\bf k}_0$. (b): calculation of
the averaged angular momentum distribution for atoms in the
periodic potential (\ref{realpotential}) with $V_0 = -4$, $V_1 =
4.664$ after the wave packet has been displaced to ${\bf k}_0$.
The averaging assumes an experimental precision in position of
$\sigma_{\rm inst} = a$. } \label{figure3.fig}
\end{figure}

The angular momentum that we have calculated is the total value
for the wave packet. To study its distribution in real space, we
consider a Gaussian distribution of Bloch waves around the point
${\bf k}_0$, i.e.
\begin{equation}
f({\bf k}) = {1\over \sqrt{\pi} \sigma_k} e^{-{({\bf k}-{\bf
k}_0)^2 \over 2 \sigma_k^2}}.
\end{equation}
In figure \ref{figure3.fig}(a) we show the
distribution of currents as a function of position, which
show a clear rotating pattern. This
distribution of currents is not the one found near a vortex, for
which the currents diverge at the center.
Approximating the off-diagonal term in the Bloch Hamiltonian near
${\bf k}_0$ as
\begin{equation}
V({\bf k}) \approx {3ta\over 2}  e^{i7\pi/6} \,((k_x - k_{0,x})
+ i(k_y - k_{0,y})),
\end{equation}
we can calculate the wave function coefficients as
\begin{eqnarray}
\psi_{m,n}^B &\approx& 2\sqrt{\pi}\sigma_k\,
e^{i{\bf k}_0\cdot{\bf R}_{m,n}} \,e^{-R_{m,n}^2\sigma_k^2/2},\\
\psi_{m,n}^A &\approx& {ta\sigma_k^2\over \epsilon_g} \,e^{i2\pi /3}
\, (R_{m,n, x} + iR_{m, n, y}) \, \psi_{m,n}^B.
\end{eqnarray}
Notice that the probability for B sites has a maximum at the
center of the wave packet, while the probability for A sites is
zero there. This is related to the fact that for a single Bloch
wave at ${\bf k}_0$ the A sites are completely depopulated, and it
is only through the presence of a distribution of Bloch waves that
some of the A sites acquire some density. The distribution of A
sites has a maximum when $R_{m,n} \approx 1/\sigma_k$. Since the
currents depend on the values at neighboring sites, it is at this
distance that the currents are maximized.

In a real experiment, the atoms will move in an optical lattice
generated by counterpropagating
laser beams or a holographic method, in which the
potential is generated with a phase or amplitude mask~\cite{mask}.
The potential to be chosen can be obtained by calculating the
Fourier transform of a sum of Gaussian wells located at the A and
B sites and keeping the largest terms. The potential obtained is
of the form
\begin{eqnarray}\label{realpotential}
V_{\rm latt} ({\bf r}) &=& V_0  [ \cos(({\bf k}_1-{\bf k}_2) \cdot
{\bf r}) +
\cos ((2{\bf k}_1+{\bf k}_2) \cdot {\bf r})  \nonumber \\
&&+ \cos ((2{\bf k}_2+{\bf k}_1) \cdot {\bf r}) ]
\nonumber \\
&&+V_1  [\cos({\bf k}_1 \cdot {\bf r}+\alpha) + \cos ({\bf k}_2
\cdot {\bf r}+\alpha) \nonumber\\
&&+ \cos (({\bf k}_1+{\bf k}_2) \cdot {\bf r}-\alpha) ].
\end{eqnarray}
Here the wavevectors are ${\bf k}_1 = -{2\pi\over 3a} {\bf e}_x
+{2\pi\over \sqrt{3}a} {\bf e}_y$, and ${\bf k}_2 = {4\pi\over 3a}
{\bf e}_x$. We will use a system of units in which $\hbar = M = a
= 1$. It is worth noticing
that the ratio between the two amplitudes $V_1$ and $V_0$ controls
the widths of the gaussian potentials located at the
lattice sites, while the phase yields the ratio between
the depths of the potentials. For the symmetric potential, $\alpha
=\pi/3$. In our calculations we have used a value of $\alpha=1.1$.

A wave packet of the form (\ref{expansion}) at ${\bf k}_c = 0$ in
the lowest energy band of a periodic potential can be obtained
starting with a Gaussian wave packet in real space (e.g. the
ground state of a harmonic trap) and adiabatically turning on the
lattice~\cite{adiabatic}. The study of the angular momentum as a
function of the quasimomentum of the wave packet can be performed
by applying an ``electric'' field (see Eq. (\ref{semiclassical2})),
which in the case of cold atoms corresponds to an acceleration of
the lattice. The atoms are displaced in reciprocal space along the
direction of the acceleration, while they perform Bloch
oscillations in real space. 

\begin{figure}
\centering \epsfig{file=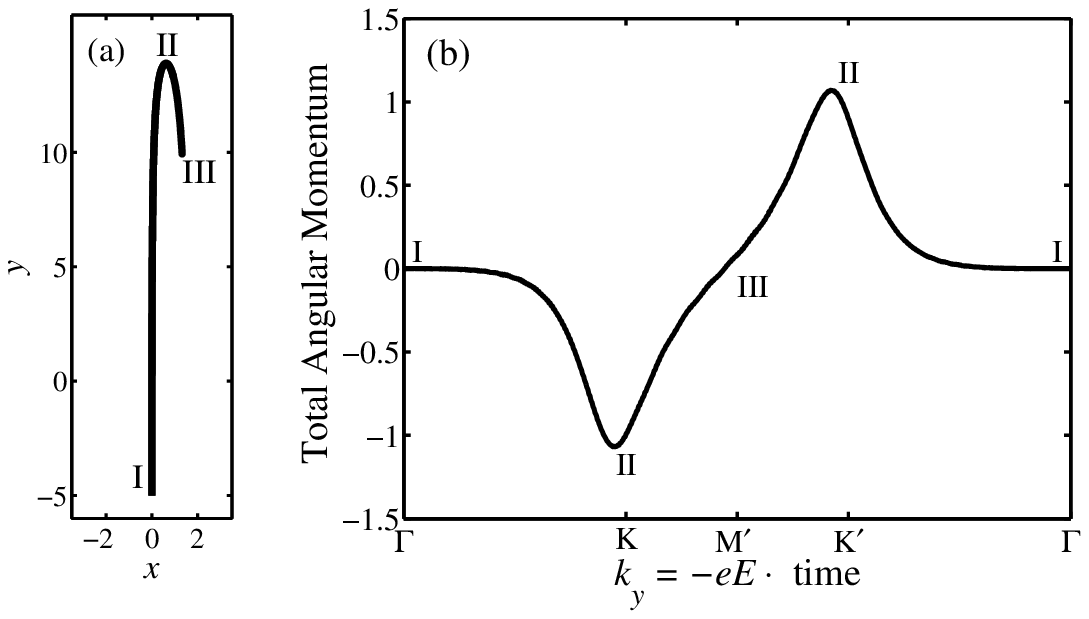, width=3.5in} \caption{
(a): motion in real space of the center of the wave packet.
(b): angular momentum as a function of the position in reciprocal 
space, calculated for an
``electric'' field of $-eE=0.05$ applied along the $y$-direction.}
\label{figure4.fig}
\end{figure}

The calculated motion of the wave packet in real space and its
angular momentum as a function of time are shown in Fig.
\ref{figure4.fig}. The wave packet starts at I, moving against the
direction of the acceleration. As the quasimomentum reaches the K
point in reciprocal space (II in real space), the Berry curvature
displaces the distribution to the right, and the angular momentum
achieves its smallest value. The averaged distribution of the
angular momentum is shown in the right panel of Fig.
\ref{figure3.fig}, which agrees with the result for the
tight-binding model (left panel of the same figure). Eventually,
the wave packet reaches point III, where it shows no rotation and
starts retracing its path. When it reaches II again, it possesses
a maximum (positive) angular momentum, which it looses once it
comes back to I.

The velocity distribution of the atoms (from which the 
angular momentum of the wave
packet can be calculated) can be measured by Doppler-sensitive Raman
transitions~\cite{Kasevich}. In this way, atoms with a selected
velocity are driven to a dark state, and one can get rid of the
rest of the atoms by shining resonant light. Imaging the remaining
atoms, one may obtain the real space distribution of atoms with the
given velocity. By changing the detuning, different sets of atoms
are observed.
Another observable result in this case is the motion caused by the
anomalous velocity (due to the nonvanishing Berry curvature). This
displaces the center of the wave packet by an amount equal to
\begin{equation}
\Delta {\bf r}_c = -\int_{\bf 0}^{{\bf k}} d{\bf k}' \times {\bf
\Omega}({\bf k'}),
\end{equation}
an amount independent of the magnitude of the driving $-eE$ field.

Care must be taken in the experiment not to excite the atoms to a
higher energy band while the lattice is accelerated. The largest
probability for this to happen occurs when the wave packet is at
${\bf k}_0$, where the gap between the two bands is smallest and
the particles can undergo Landau-Zener tunnelling. The probability
for this to happen is proportional to $\exp(-\epsilon_{g}^2/|eE|)$,
therefore the electric field must satisfy $|eE| \ll \epsilon_g^2$.

One has to be careful in the design of the lattice in which the
rotation is observed in order to make a direct comparison with the
tight-binding model. Using a potential depth $V_0$ much smaller
than the one we used in the simulations one has to include
the probability of atoms tunnelling to a non-nearest neighbor. 
These currents
between B sites would not add to the total angular momentum in the
wave packet but can yield a much more complicated distribution of
currents.

\begin{figure}
\centering \epsfig{file=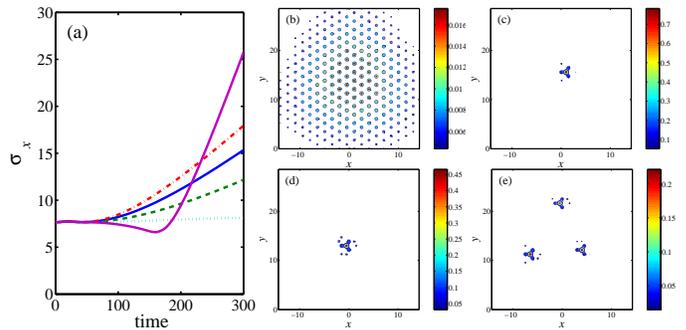, width=3.5in}
\caption{(a) Dispersion of the wave packet with time for different
values of the interaction strength in the tight-binding limit 
($\epsilon_g = 6$, $t=1$). The wave packet is accelerated from the 
$\Gamma$ point
to the $K$ point, where the interaction strength is adiabatically
increased. The lines correspond to $g=0$ (blue line),
$g=-1$ (red line), $g=1$ (green line), $g=1.9$ (cyan line), and
$g=3$ (magenta line). In this last case, part of the wave packet
collapses to a soliton when the size of the wave packet is
minimum, with the remaining part diffusing at a fast rate
afterwards. (b)-(e) show the wave function 
in real space
after dispersing in the lattice for
$t \approx 200$; the interaction strengths are 
$g=1.5$ (b), $g=7.5$ (c), $g=15$ (d),
and $g=30$ (e).} \label{figure5.fig}
\end{figure}

We finally comment on the consequences of adding the interatomic
interaction between the particles.  The interaction corresponds to
the addition of terms $g|\psi_{m,n}|^2 \psi_{m,n}$ in the right
hand sides of (\ref{psiA}) and (\ref{psiB}). For weak interaction
strength we observe no change in the values of the angular
momentum at ${\bf k}_0$, which remains constant in time. The rate
of dispersion of the wave packet is modified, with a repulsive
interaction ($g>0$) slowing down the dispersion and an attractive
interaction ($g<0$) accelerating it (see Fig. \ref{figure5.fig}).
This counterintuitive result is related to the negative effective 
mass
associated with particles near the top of the energy
band~\cite{Meystre}. Moreover, for repulsive interactions larger
than a critical value, the system exhibits collapse into one or
more breathing solitons~\cite{quasicollapse}. These can be related
to solitons in the gap between the two bands~\cite{Kivshar}
similar to recently observed two dimensional optical
solitons~\cite{Segev} and will be considered in a separate
article.

The authors would like to acknowledge discussions with Mark G. Raizen. 

%


\begin{thebibliography}{99}
\bibitem{adiabatic}
The condition to remain in the ground state can be
estimated as $t_{\rm turn \, on} \gg 1/\Delta \cal E$, where
$\Delta \cal E$ is the smallest excitation energy
at ${\bf k}=0$ as the potential is turned on.
\end{thebibliography}
\end{document}